\documentclass[fleqn,12pt]{wlscirep}
\usepackage[utf8]{inputenc}
\usepackage[T1]{fontenc}
\usepackage{grffile,float,xcolor,xspace}
\usepackage{amssymb,amsfonts,amsmath,graphicx}
\usepackage{subfigure}
\title{Enhancement of Extreme Events through the Allee effect and its Mitigation through Noise in a Three Species System}
\author[1]{Deeptajyoti Sen}
\author[1,*]{Sudeshna Sinha}
\affil[1]{Indian Institute of Science Education and Research Mohali,\\
Knowledge City, Sector 81, Manauli PO 140306, India\\
\affil[*]{sudeshna@iisermohali.ac.in}}

\begin{abstract}
We consider the dynamics of a three-species system incorporating the Allee Effect, focussing on its influence on the emergence of extreme events in the system. First we find that under Allee effect the regular periodic dynamics changes to chaotic. Further, we find that the system exhibits unbounded growth in the vegetation population after a critical value of the Allee parameter. The most significant finding is the observation of a critical Allee parameter beyond which the probability of obtaining extreme events becomes non-zero for all three population densities. Though the emergence of extreme events in the predator population is not affected much by the Allee effect, the prey population shows a sharp increase in the probability of obtaining extreme events after a threshold value of the Allee parameter, and the vegetation population also yields extreme events for sufficiently strong Allee effect. Lastly we consider the influence of additive noise on extreme events. First, we find that noise tames the unbounded vegetation growth induced by Allee effect. More interestingly, we demonstrate that stochasticity drastically diminishes the probability of extreme events in all three populations. In fact for sufficiently high noise, we do not observe any more extreme events in the system. This suggests that noise can mitigate extreme events, and has potentially important bearing on the observability of extreme events in naturally occurring systems.
 
\end{abstract}

\begin{document}

\flushbottom
\maketitle

\section{Introduction}

The emergence of extreme events in the dynamical evolution of systems ranging from  weather \cite{weather} to power grids \cite{powergrid1,powergrid2} have catastrophic implications. So understanding the underlying mechanisms that may trigger extreme events have commanded considerable recent research interest \cite{extreme}. An extreme event may be defined as an event where one or more variables of a system, arising in nature or in the laboratory, exhibits very large deviations from the mean value. So the dynamics of the system is characterized by excursions to values that differ significantly from the average. Further, though recurrent, these large deviations are rare vis-a-vis the characteristic time scale of the system, and their occurrences are aperiodic and uncorrelated in time. Without loss of generality, an event is typically labelled `extreme' if a state variable, in the course of its temporal evolution, takes values that are several standard deviations away from the average value, thereby signalling dynamical behaviour beyond normal variability. Such extreme events have been observed in natural systems such as rogue ocean waves \cite{ocean}, laboratory systems such as optical systems \cite{optical}, as well as financial phenomena like market crashes \cite{market}. Interestingly, other definitions of extreme events, more appropriate to the context, have also been employed. Notably, in the specific important problem relating climate change to ecological dynamics, a synthetic definition of extreme events involving both the driver and response system is proposed \cite{extreme_new_def}. However in this work we will employ the most commonly used marker for extreme events: namely, recurrent and aperiodic deviations larger than a prescribed threshold from the mean value, are considered to be extreme events, with the threshold typically taken to be 3-8 standard deviations from the mean.

A central direction in understanding extreme events is to find generic mechanisms that can give rise such large fluctuations. Typical studies of extreme events have involved stochastic models \cite{satya1,satya2}, for instance random walk models of transport on networks \cite{santhanam}. The emergence of extreme events in deterministic dynamical systems, manifested as intermittent large amplitude events, have also been investigated recently. Broadly speaking, the statistical features of deterministic systems is an active research direction that can lead to understanding extreme events that arise in the context of deterministic dynamics \cite{balki1,balki2}. The search for dynamical systems that yield extreme events, without the drive of external stochastic influences or intrinsic random fluctuations, is a focus of much ongoing research effort from the point of view of basic understanding of complex systems  \cite{ulrike1,ulrike2,ulrike3,promit}. Additionally, such probabilistic outcomes in dynamical systems are most relevant in the applied context as well, such as in engineering sciences where this direction of research leads to better assessment of risks \cite{sujith}.

In this broad direction, in this work we explore the dynamics of a vegetation-prey-predator system, coupled through interactions of the Lotka-Volterra type. Importantly, our model system also incorporates the biologically significant Allee effect, which is one of the classic phenomena of population ecology. The Allee effect reflects the beneficial effects on the growth of individuals arising from conspecific interactions \cite{Courchamp2008,Sen2021,Dennis1989}, and has bearing on the long-term persistence of a population as a consequence of small size. Further we investigate the effect of additive noise in the system, focusing on the role of stochasticity in the emergence of extreme events.
In a larger context, this research direction also has bearing on the important general question of the emergence of extremely large events in deterministic dynamical systems, and the effect of noise on their sustained prevalence.

Our central findings in this work are as follows: the system under Allee effect yields chaotic dynamics. Further, it leads to unbounded vegetation growth for sufficiently strong Allee effect. Most significantly, Allee effect aids the emergence of extreme events in this three-species chain. Interestingly, we also observe that noise suppresses the unbounded blow-up of vegetation induced by Allee effect. Lastly, sufficiently strong noise also subdues the extreme events in vegetation, prey and predator populations, thus suggesting a significant natural mechanism to mitigate extreme events in population chains.

In Section 2 we present results arising in the model of three interacting species, incorporating the Allee effect. In Section 3 we explore the effect of additive noise in the system. We conclude with discussions of the scope of our findings in Section 4.

\section{Three-species food chain model incorporating the  Allee Effect}
 
 Complex systems research in general, and theoretical ecology in particular, has seen intense research activity in networks modelling interacting species, often focusing on local and global stability properties \cite{blasius,maywigner,robust,balance}. Here we will focus on the emergence of extreme events and consider as our test-bed the well-known model for the dynamics of the snowshoe hare and the Canadian lynx populations, based on observed data \cite{blasius}. Specifically, the system incorporates a three species vertical food chain, consisting of vegetation (denoted by $u$), prey (denoted by $v$) and predator (denoted by $w$). Additionally, we will incorporate in this model a term to reflect the Allee effect in the growth of the prey. The dynamics of this three species trophic system is described by the following sets of coupled differential equations:
 
 \begin{equation}
 \label{eq:finalmodel}
     \begin{split}
         \dot{u} &= f(u,v,w)\,=\,a u \ - \ \alpha_{1}f_{1}(u,v),\\
         \dot{v} &= g(u,v,w)\,=\,\alpha_{1}f_{1}(u,v) \ A(v) \ - \ bv \ - \ \alpha_{2}f_{2}(v,w),\\
         \dot{w} &= h(u,v,w)\,=\,\alpha_{2}f_{2}(v,w) \ - \ c(w-w^{*}).
     \end{split}
 \end{equation}

 \noindent The interaction between the vegetation and prey populations is considered to follow the type II functional response, described by the function $f_{1}(u,v)$, where\\ $f_{1}(u,v)\ = \ \frac{u v}{1 + k u}$. This well-known functional response is characterized by a decelerating intake rate, stemming from the assumption that the consumer is limited by its capacity to process food. The parameter $k$ is the average time spent on processing a food item, which is termed the handling time in literature.

The interaction of the predator population with the prey is considered to follow the well-known Lotka-Volterra type interaction, described by the function $f_{2}(v,w)$, where $f_{2}(v,w)\ = \ v  w$. Here $\alpha_{1}$ denotes the maximum growth rate of the prey, which is in general a product of ingestion rate with a constant factor ($<1 $), accounting the fact that not all of the ingested resource (vegetation) converted into prey's biomass. A similar parameter for the predator is denoted by $\alpha_{2}$. The parameters $a$, $b$ and $c$ represent the intrinsic growth rates of the three species $u$, $v$ and $w$ respectively. Further, the model allows the predator population to maintain a equilibrium population $w^{*}$ when the prey concentration is very low. In other words, predator can survive in the trophic system without depending on prey. 

Importantly, in contrast to work on similar systems \cite{Chaurasia2020}, here we will also explicitly consider the Allee effect. Specifically, the Allee effect is considered in growth of the prey by introducing  
\begin{equation}
    A(v)\,=\,\frac{v}{v+\theta},
\end{equation}
\noindent where $\theta$ is the Allee strength parameter representing the critical prey density at which the probability of successful mating would be half. The Allee effect here occurs due to difficulties in finding mates for sexual reproduction and $A(v)$ describes the mating success at the low population density \cite{Boukal2002,Rowe2004}. This kind of Allee effect in two and three dimensional population models has drawn much research attention \cite{Boukal2002,Courchamp2008,Sen2021,Dennis1989,Mccarthy1997,Scheuring1999}. 

In this work we consider the parameter values  $a\,=\,1$, $b\,=\,1$, $c\,=\,10$, $w^{*}\,=\,0.006$, $\alpha_{1}\,=\,0.5$, $\alpha_{2}\,=\,1$, $k\,=\,0.05$ \cite{blasius}. We explore the dynamics of the system under varying $\theta$, through numerical simulations using the Runge-Kutta fourth order algorithm. We have ascertained the stability and convergence of our results with respect to decreasing step size.

Our main focus  is to explore the dynamical consequences of the Allee effect in the population of the prey, in this three-species model system. We will first show that there is a sharp increase in the probability of obtaining unbounded vegetation growth beyond a critical value of the Allee parameter. We will then demonstrate that, interestingly, the system exhibits chaos as the Allee effect becomes more significant. Then we will move on to the central focus of this work, namely the influence of Allee effect on the emergence of extreme events.  We will explicitly demonstrate that there is a pronounced increase in the propensity of extreme events under increasing Allee parameter $\theta$, and also show that the extreme events occur with strikingly different probabilities for the different species.


\section*{Temporal evolution of the population densities}

\begin{figure}
    \centering
    \includegraphics[width = 0.55\textwidth]{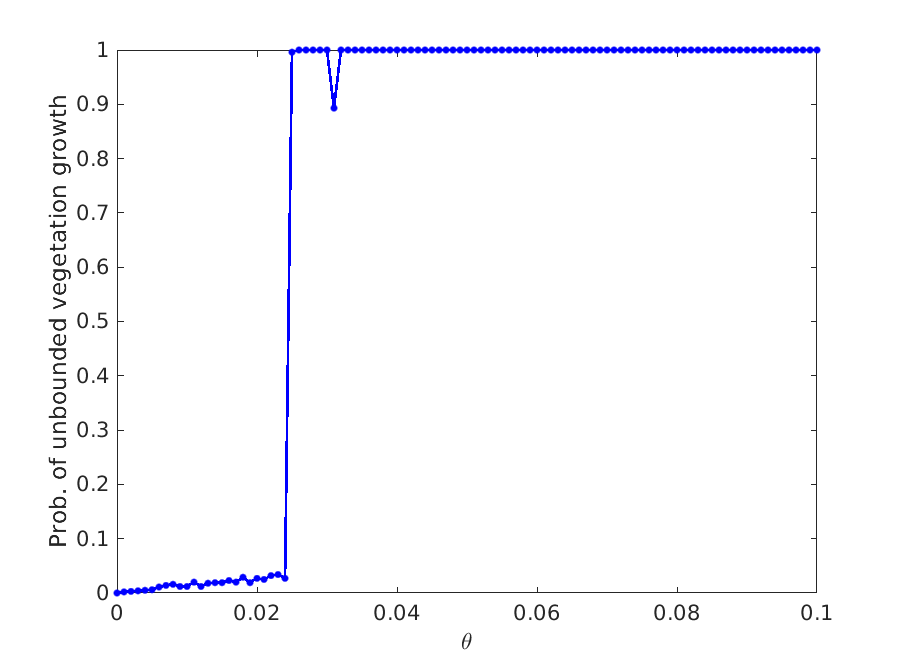}
    \caption{Probability of unbounded growth of the vegetation population ($u$) with respect to $\theta$. Here an explosive blow-up is considered to have occurred when the vegetation population exceeds a value of  $10^3$. The probability is estimated from a sample of $10^3$ initial states randomly distributed in a hyper-cube ($u \in [0:4], v \in [0:2], w \in [0:5]$) in phase space. Interestingly, the slight dip in the estimated probability of unbounded growth in the narrow window around $\theta \sim 0.031$ arises from unbounded orbits co-existing with a small set of initial states that evolve to bounded periodic orbits.}
    \label{fig:Prob_blowup}
\end{figure}

Our first observation is the emergence of explosive runaway growth in vegetation when the Allee effect is too strong, i.e. when the Allee parameter $\theta$ is sufficiently large, the vegetation grows in an unbounded manner \cite{scirep_unbounded}. In order to quantify this blow-up, we estimate the probability of unbounded vegetation growth
from a large sample of random initial states, followed over a long period of time. We also ascertain that the estimated values are converged with respect to increasing sample size, and thus can be considered to be robust numerically. The results thus obtained, for varying Allee parameter $\theta$, are displayed in Fig.~\ref{fig:Prob_blowup}.  It is clearly evident from the figure that 
there exists a critical value of $\theta$, which we denote by $\theta_{c}$, beyond which the vegetation has a probability of explosive unbounded growth. So in the rest of this work, we will restrict our analysis for the range $\theta \in [0,\theta_{c})$. 

\begin{figure*}
\centering
\mbox{\subfigure[]{\includegraphics[width=0.48\textwidth]{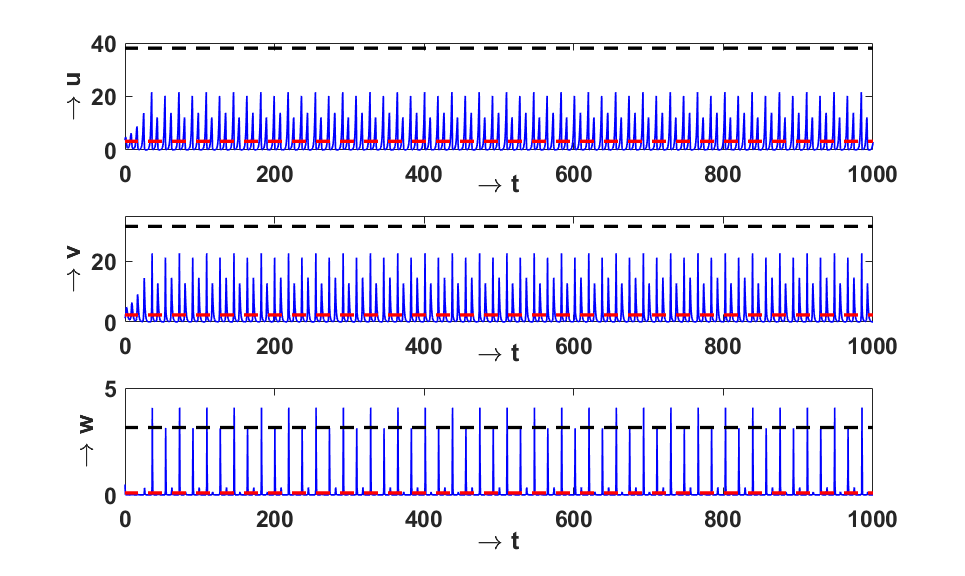}}
\quad
\subfigure[]{\includegraphics[width=0.48\textwidth]{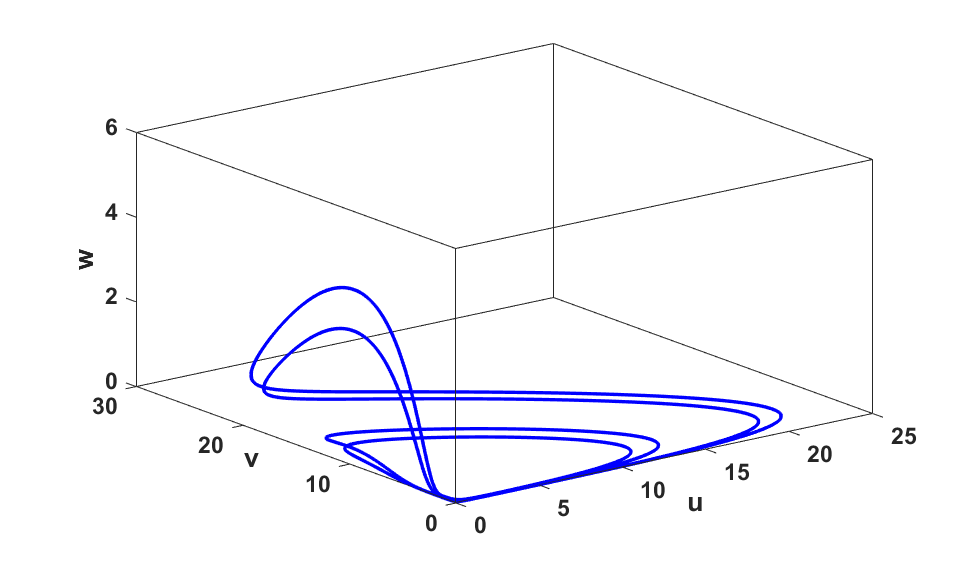}}}
\mbox{\subfigure[]{\includegraphics[width=0.48\textwidth]{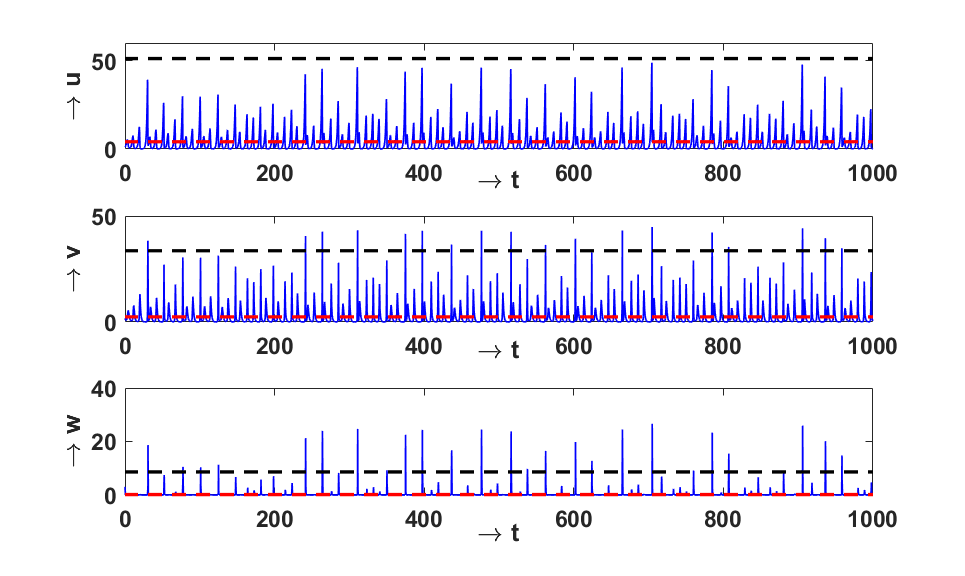}}
\quad
\subfigure[]{\includegraphics[width=0.48\textwidth]{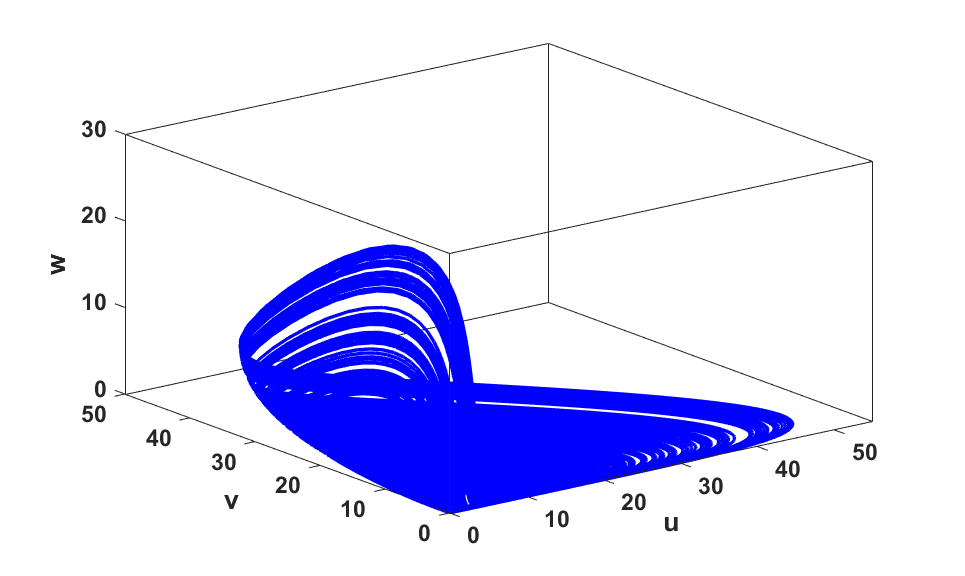}}}
\mbox{\subfigure[]{\includegraphics[width=0.48\textwidth]{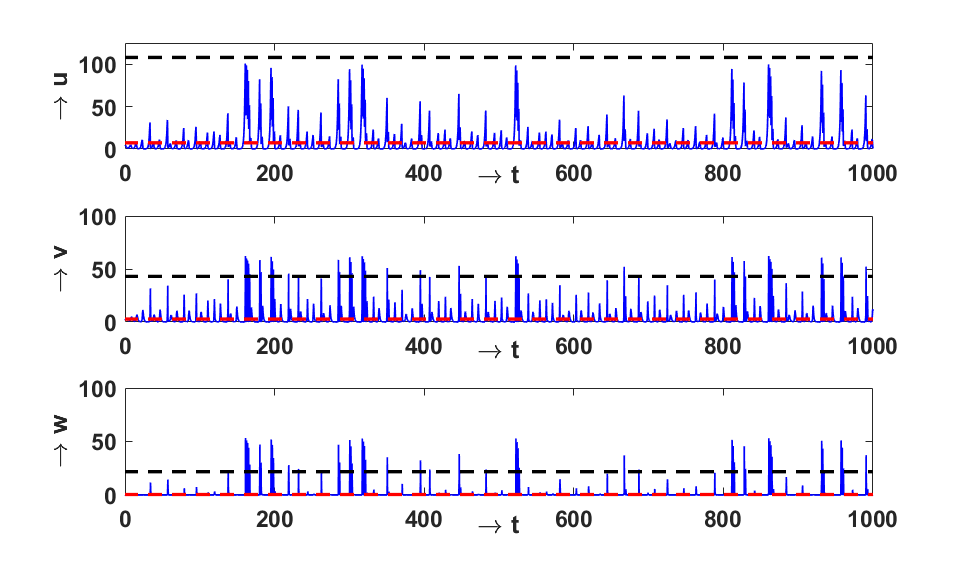}}
\quad
\subfigure[]{\includegraphics[width=0.48\textwidth]{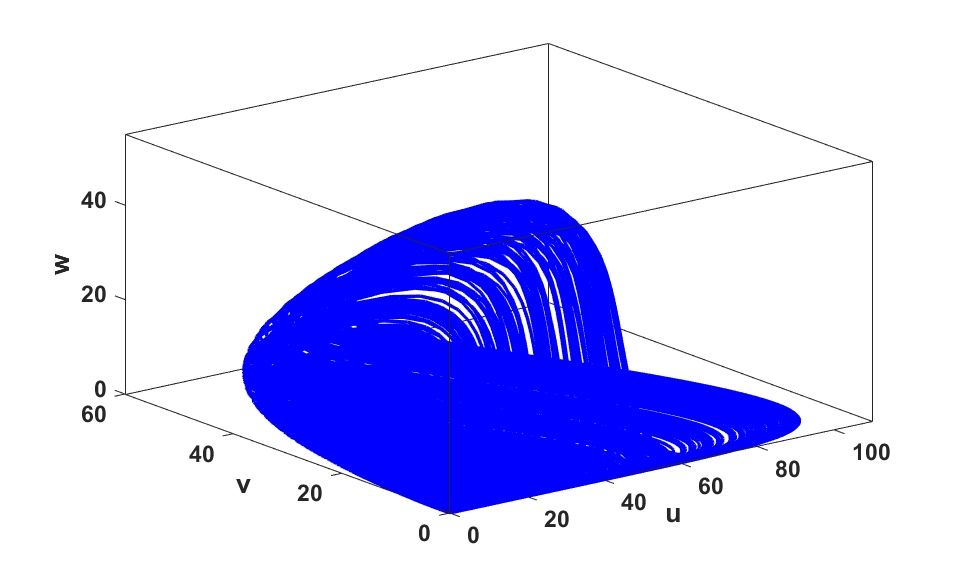}}}
\caption{
Left panels display the time series for the vegetation ($u$), prey ($v$) and predator ($w$) populations in the system given by Eqns.~\ref{eq:finalmodel}, and the right panels display the corresponding phase space attractor. The Allee effect parameter is $\theta\,=\,0$ (a-b), $\theta\,=\,0.01776$ (c-d), and $\theta\,=\,0.02475$ (e-f). The red dashed line shows the mean $\mu$, and the black dashed line represents the threshold level of $7$ standard deviations above the mean (i.e. $\mu + 7\sigma$).}
\label{fig:timeseries&phasep}
\end{figure*}

\begin{figure}
    \centering
    \includegraphics[width=0.6\textwidth]{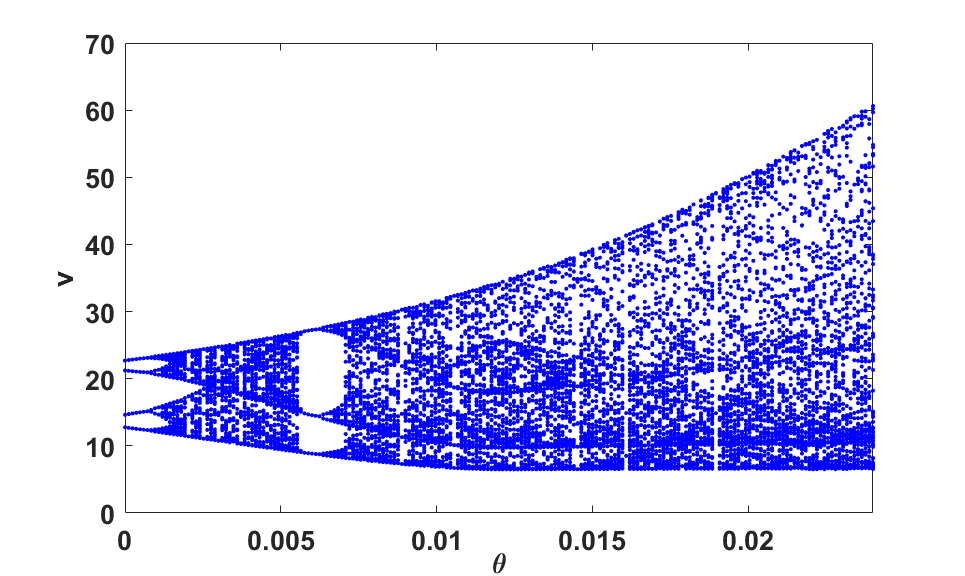}
    \caption{Bifurcation diagram of prey populations with respect to Allee parameter $\theta$. Here we display the local maxima of the prey population. 
The parameter values in Eqn.~\ref{eq:finalmodel} are as mentioned in the text.}
    \label{fig:Bfdiag_wrt_theta}
\end{figure}

Next we investigate the temporal evolution of the population densities. Fig.~\ref{fig:timeseries&phasep} shows representative time series of vegetation, prey and predator and the corresponding attractors in 3-dimensional phase space. To broadly illustrate the influence of Allee effect, we present this
for three values of $\theta$, with increasing magnitude. When Allee effect is absent in the prey population i.e. $\theta\,=\,0$, we observe that all populations fluctuate periodically and are confined to a periodic orbit [see Fig.~\ref{fig:timeseries&phasep}a and Fig.~\ref{fig:timeseries&phasep}b]. For a larger Allee parameter ($\theta\,=\,0.01776$), the populations of vegetation, prey and predator all evolve in an aperiodic manner, as evident in Fig.~\ref{fig:timeseries&phasep}c, with the corresponding chaotic attractor shown in Fig.~\ref{fig:timeseries&phasep}d. 
On further increasing $\theta$ to $0.02475$, the size of the chaotic attractor increases, as evident from Fig.~\ref{fig:timeseries&phasep}e,f. Therefore increasing magnitude of the Allee effect parameter drives the system into a chaotic state from the periodic state. To corroborate this observation, we present the bifurcation diagram of the prey population, with respect to the Allee parameter $\theta$, in Fig.~\ref{fig:Bfdiag_wrt_theta}. We observe the onset of chaos through the usual period-doubling cascade, initiating at period-$4$ at $\theta=0$. Subsequently, we also observe narrow periodic windows, with period-$3$ being the most prominent one. The most significant implication of this bifurcation diagram is the emergence of chaotic dynamics in a very large range of the Allee parameter $\theta$. That is, with no Allee effect or under very weak Allee effect the system is periodic, while a {\em strong Allee effect typically induces chaos in this three-species system.}  


\section*{Extreme events induced by Allee Effect}

One of the most interesting observations from the time series presented in the section above is the following: when the magnitude of the Allee parameter $\theta$ is low, vegetation and prey densities are confined to low values. However, the predator densities deviate very significantly away from their mean. Now for very small $\theta$ the system is attracted to a periodic orbit, and so the large deviations are completely correlated with time and occur periodically. So they cannot be considered to be extreme events, as they are neither aperiodic, nor rare. But for larger $\theta$, both predator and prey densities can sometime shoot up over 7 standard deviations away from the mean value. This is evident clearly in Fig.~\ref{fig:timeseries&phasep}(c) \& (e) where one can see that both predator and prey populations exceed the $7\sigma$ threshold from time to time. The instants at which prey and predator populations exceed the $7\sigma$ threshold are now completely uncorrelated with time. This is consistent with the underlying chaotic dynamics that emerges under increasing Allee parameter $\theta$. 

\begin{figure}
    \centering
    \includegraphics[width=0.65\textwidth]{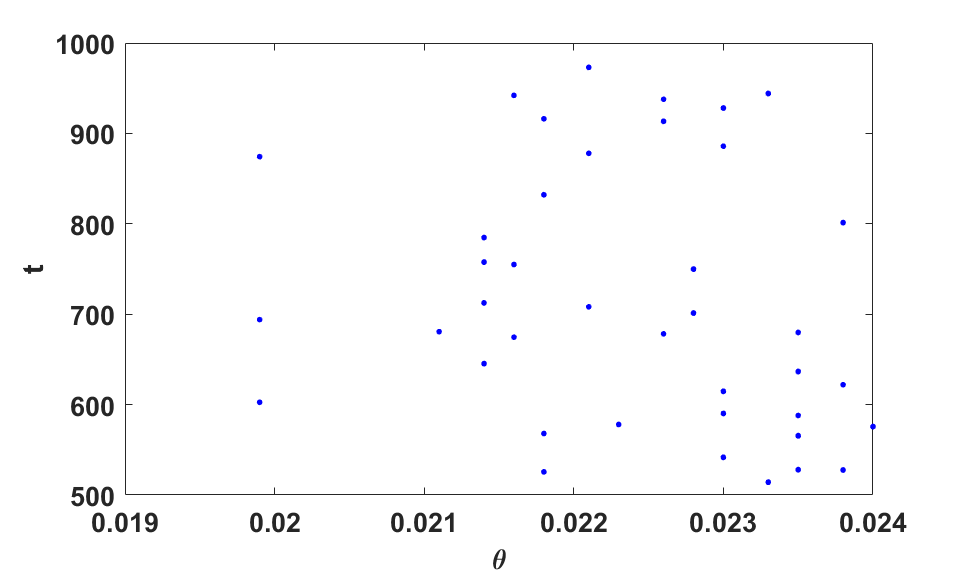}
\includegraphics[width=0.65\textwidth]{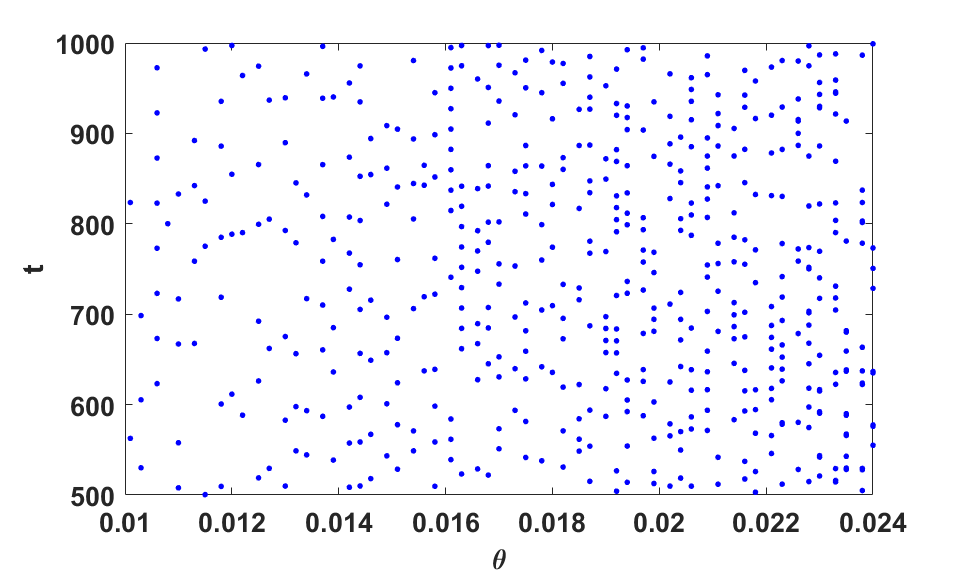}    
    \includegraphics[width=0.65\textwidth]{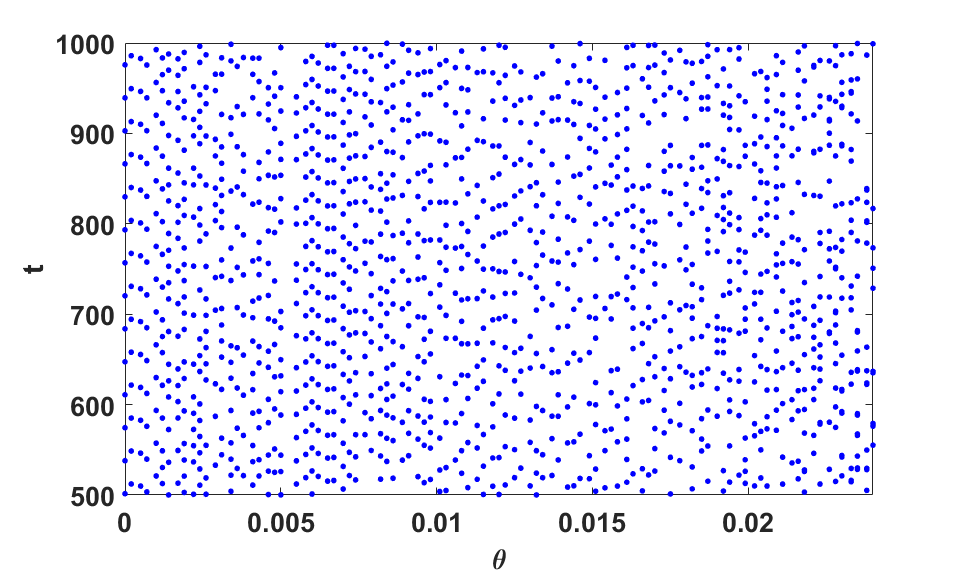}
    \caption{Figure marking the time instances at which a population exceeds the $7\sigma$ threshold, for different values of Allee parameter $\theta$, for the  case of (top to bottom) vegetation, prey and predator populations. }
    \label{t_extreme}
\end{figure}

In order to illustrate this, we mark the time instances at which a population exceeds the $7\sigma$ threshold, for different values of Allee parameter $\theta$. Fig.~\ref{t_extreme} shows this for the vegetation, prey and predator populations. The density of points signifying the occurrence of extreme events is clearly the highest for the predator population. This indicates that the predator population has the greatest propensity for large deviations. It is also clear that vegetation has the least number of extreme events in the same time window. The uncorrelated nature of the extreme events is also evident in the scatter of these points, except in the small periodic windows that occur for certain special ranges of $\theta$. The increasing density of these points also illustrate the increasing probability of extreme events in the populations with increasing Allee parameter $\theta$.


In order to understand the phenomena quantitatively, we first estimate the maximum densities of vegetation, prey and predator populations (denoted by $u_{max}$, $v_{max}$ and $w_{max}$ respectively) for varying the Allee parameter $\theta$. To estimate this, we find the global maximum of the populations sampled over a time interval $T\,=\,1000$, averaged over a large set of random initial conditions.

\begin{figure}
    \centering
    \includegraphics[width=0.55\textwidth]{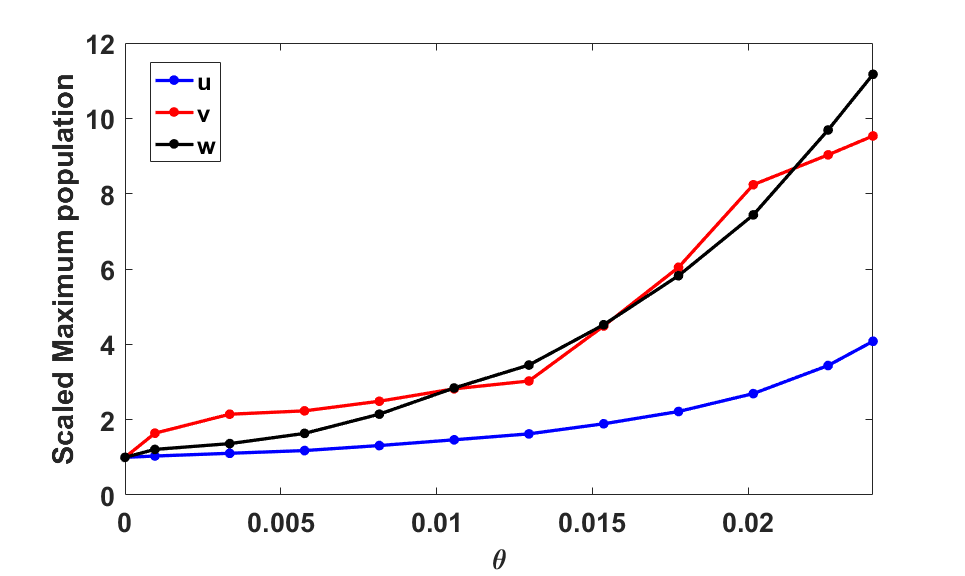}
    \caption{Global maximum of vegetation $u_{max}$ (blue),   prey $v_{max}$ (red) and predator (black) populations, with respect to the Allee parameter $\theta$, 
    scaled by their  values obtained for $\theta\,=\,0$. Clearly, when Allee parameter $\theta$ is sufficiently large, the maximum prey and predator populations are an order of magnitude larger than that obtained in systems with no Allee effect. }
    \label{fig:max_population}
\end{figure}

Fig.~\ref{fig:max_population} shows $u_{max}$, $v_{max}$ and $w_{max}$, for Allee parameter $\theta \in [0,\theta_{c})$, scaled by their values at $\theta = 0$. 
These scaled maxima help us gauge the relative change in the maximum population densities arising due to the Allee effect. It is evident from our simulation results that the magnitude of the global maximum of vegetation does not change very significantly for increasing Allee parameter $\theta$, with its magnitude around $\theta_c$ being approximately 4 fold the value at $\theta\,=\,0$. However, the magnitude of maximum prey and predator populations change very significantly with respect to Allee parameter $\theta$ and exceeds over $10$ fold the value obtained for $\theta\,=\,0$.

\begin{figure}
\begin{center}
\includegraphics[width=0.6\textwidth]{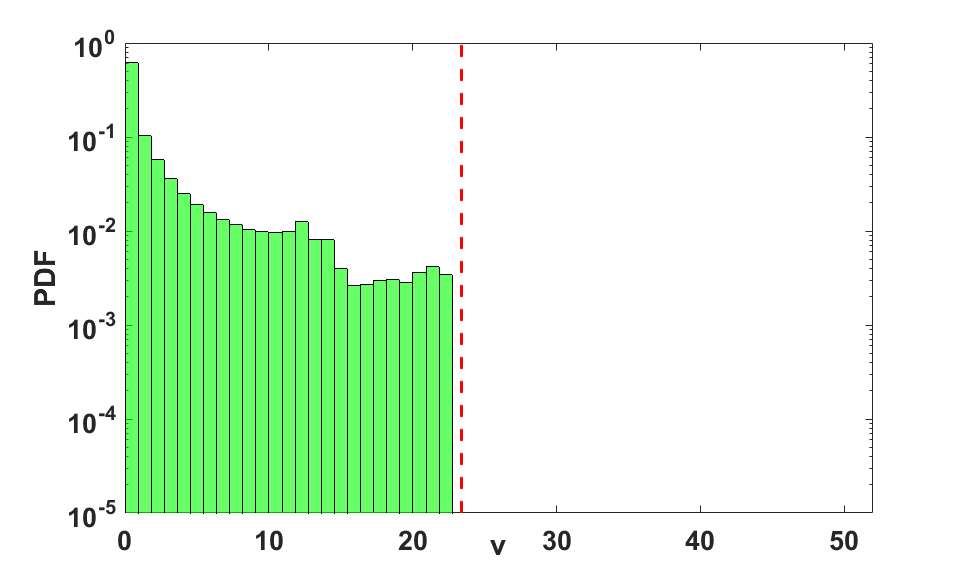}
\includegraphics[width=0.6\textwidth]{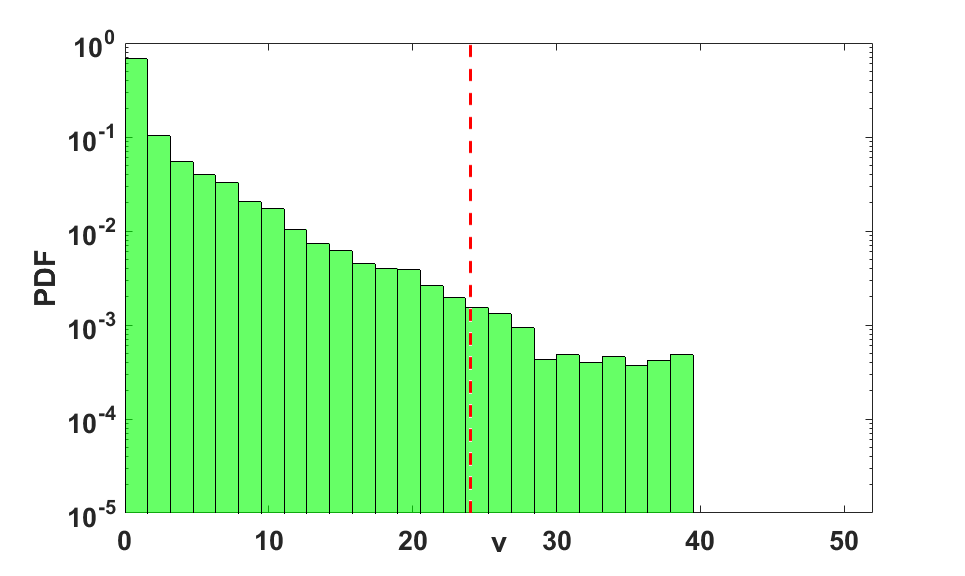}
\includegraphics[width=0.6\textwidth]{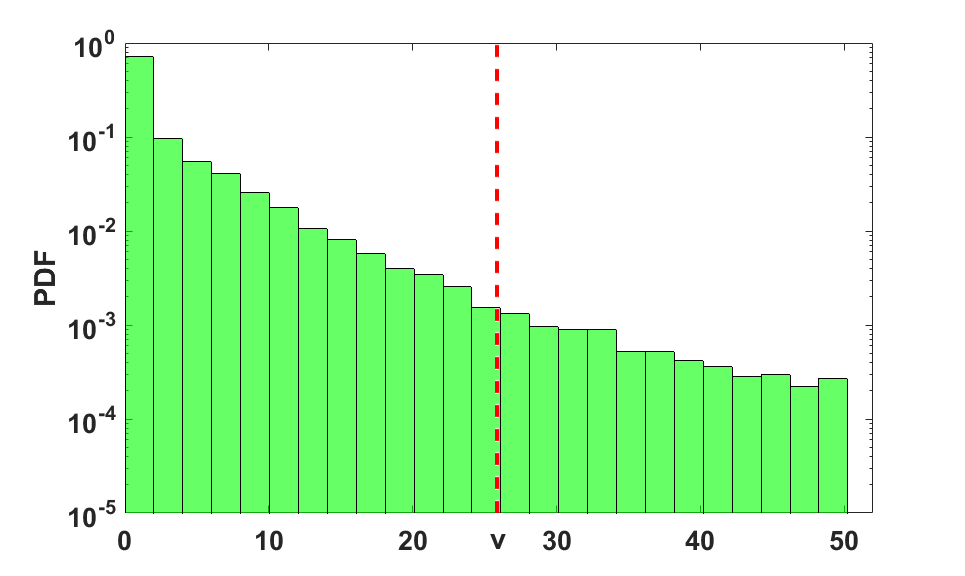}
\caption{Probability Density Function (PDF) of the prey population $v$, for the system given by Eqs.~\eqref{eq:finalmodel}, with increasing magnitude of $\theta$ with (a) $\theta\,=\,0$, (b) $\theta\,=\,0.015$ and (c) $\theta\,=\,0.02$. The threshold for extreme event $\mu + 7\sigma$ is denoted by vertical red dashed line.}
\label{fig:ProbabilityDist_population}
\end{center}
\end{figure}

We then go on to numerically calculate the probability density of the vegetation, prey, and predator population densities, for increasing Allee effect parameter $\theta$. The tail of this probability density function reflects the influence of the Allee effect on the probability of obtaining extreme events. To illustrate this, we show the probability density function for the prey population in Fig.~\ref{fig:ProbabilityDist_population}, for three different values of $\theta$. Extreme events are confined to the tail of the distribution that lie beyond the vertical red line, marking the $\mu + 7 \sigma$ value in the figure. So it is clear from these probability distributions that the Allee effect in prey population promotes the occurrence of extreme events as the tail of the distribution is flatter and extends further with increasing Allee parameter $\theta$.

\begin{figure}[htb]
	\centering
	\includegraphics[width=0.49\linewidth]{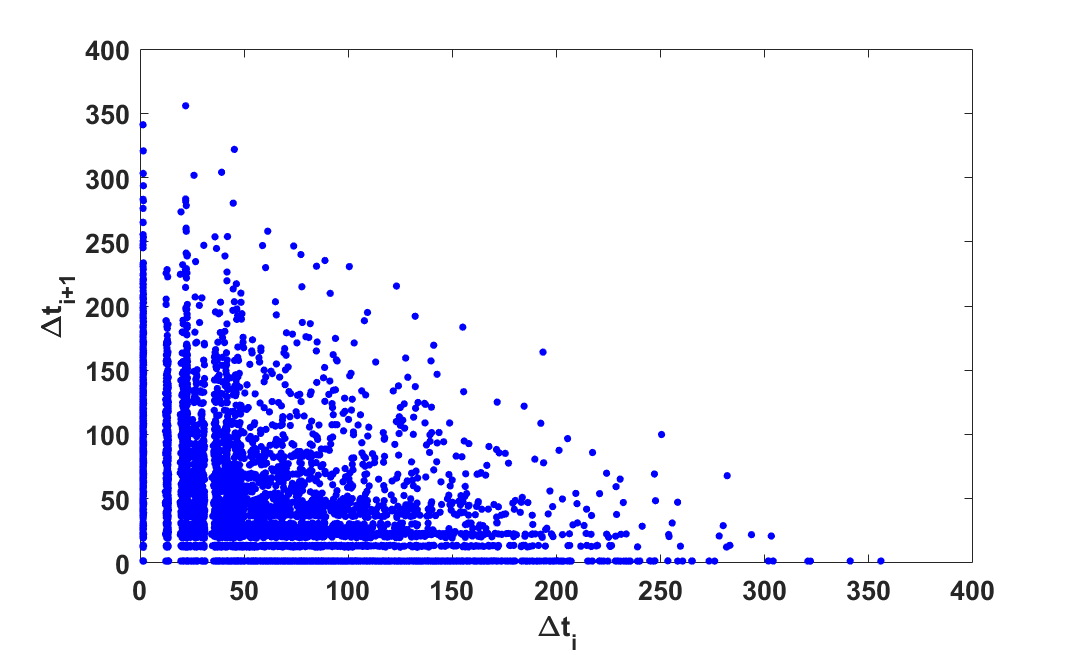}
	\includegraphics[width=0.49\linewidth]{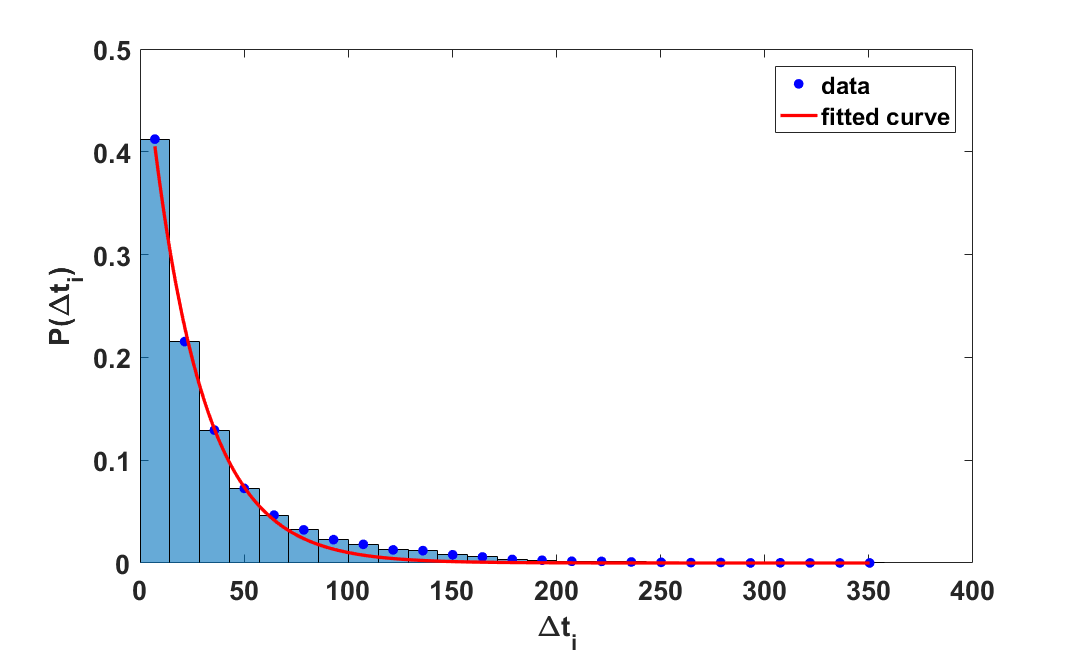} 
	\caption{(Left) Return Map of $\Delta t_{i+1}$ vs $\Delta t_i$, and (right) Probability distribution of $\Delta t_i$ fitted with exponentially decaying function, where $\Delta t_i$ is the $i^{th}$ interval between successive extreme events, where an extreme event is defined at the instant when the prey population crosses the $\mu+7\sigma$ line (cf. Fig~\ref{fig:timeseries&phasep}). Here $\theta=0.024$.}
	\label{Del_t}
\end{figure}

\begin{figure}
\centering
\includegraphics[width=0.55\textwidth]{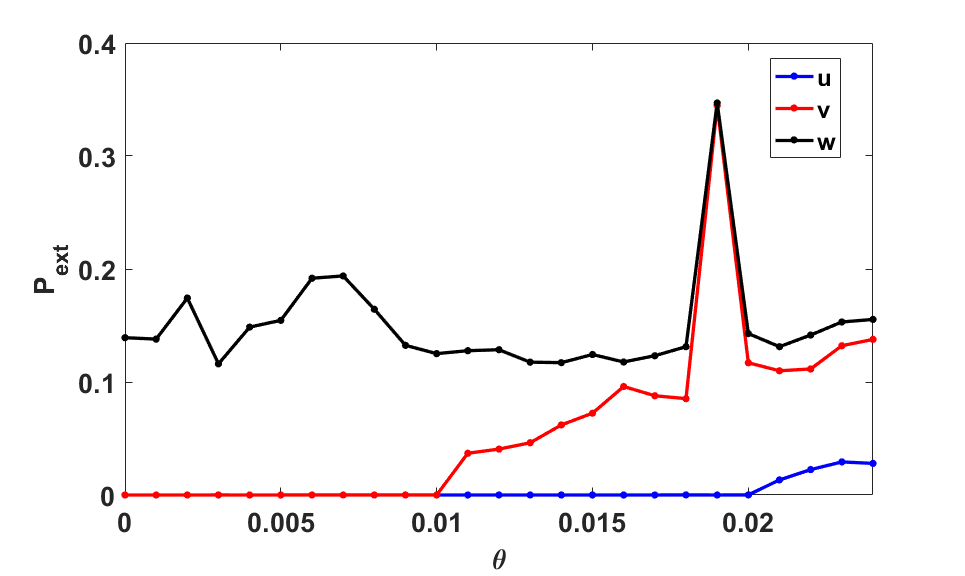}
\caption{Probability of obtaining extreme event  in unit time ($P_{ext}$), with respect to Allee parameter $\theta$, estimated by sampling a time series of length $T=5000$, and averaging over $500$ random initial states. Here we consider that an extreme event occurs when a population level crosses the threshold $\mu + 7\sigma$. $P_{ext}$ for vegetation, prey and predator are displayed in blue, red and black colors respectively. 
Note that there exists a narrow periodic window around $\theta \sim 0.02$ (cf. Fig.~\ref{zoom}), and so the large deviations in this window of Allee parameter are not associated with true extreme events, as they occur periodically. 
}
\label{fig:Dsit_no_extreme}
\end{figure}

\bigskip

In order to ascertain that the extreme values are uncorrelated and aperiodic we examine the time intervals between successive extreme events in the population. Fig.~\ref{Del_t} (left panel) shows representative results for the return map of the intervals between extreme events in the prey population  and it is clearly shows no regularity. The probability distribution of the intervals is also Poisson distributed and so the extreme population buildups are uncorrelated aperiodic events, as clearly evident from the right panel of the figure.

In order to further quantify how Allee effect influences extreme events, we estimate the probability of obtaining large deviations, in a large sample of initial states tracked over a long period of time. We denote this probability by $P_{ext}$, and we calculate it by following a large
set of random initial conditions and recording the number of occurrences of the population crossing the threshold value in a prescribed period of time, with this time window being several orders of magnitude larger than the mean oscillation period. This time-averaged and ensemble-averaged quantity yields a good estimate of $P_{ext}$. With no loss of generality, we choose the threshold for determining extreme events to be $\mu + 7 \sigma$, i.e. when the variable crosses the $7 \sigma$ level, it is labelled as extreme.

This probability, estimated for all three populations is shown in Fig.~\ref{fig:Dsit_no_extreme}.  First, it is clear from Fig.~\ref{fig:Dsit_no_extreme}, that the probability of the occurrence of extreme events is the lowest for vegetation, and the highest for predator populations, for any value of the Allee parameter $\theta \in [0,\theta_{c})$.  We also observe that, for values of the Allee parameter $\theta$ lower than a critical value denoted by $\theta^{u}_{c}$ the probability of obtaining extreme events in the vegetation population tends to zero. Beyond the critical value $\theta^u_c$, the vegetation population starts to exhibit extreme events. A similar trend emerges for the prey population. However, the critical value of the Allee parameter $\theta$ necessary for the emergence of a finite probability of extreme events, denoted by $\theta^{v}_{c}$, is much smaller than $\theta^u_c$. So for the prey population, a weaker Allee effect can induce extreme events.

Note that some mechanisms have been proposed for the generation of extreme events in deterministic dynamical systems, which typically have been excitable systems. These include interior crisis, Pomeau-Manneville intermittency, and the breakdown of quasiperiodic motion. However the extreme events generated by these mechanisms occur typically at very specific critical points in parameter space, or narrow windows around it. The first important difference in our system here is that the extreme events do not emerge only at some special values alone. Rather, there is a broad range in Allee parameter space where extreme events have a very significant presence. This makes our extreme event phenomenon more robust, and thus increases its potential observability. This also rules out the intermittency-induced mechanisms that have been proposed, as is evident through the lack of sudden expansion in attractor size in our bifurcation diagram (Fig.~\ref{fig:Bfdiag_wrt_theta}) in general.

\begin{figure}
    \centering
    \includegraphics[width=0.6\textwidth]{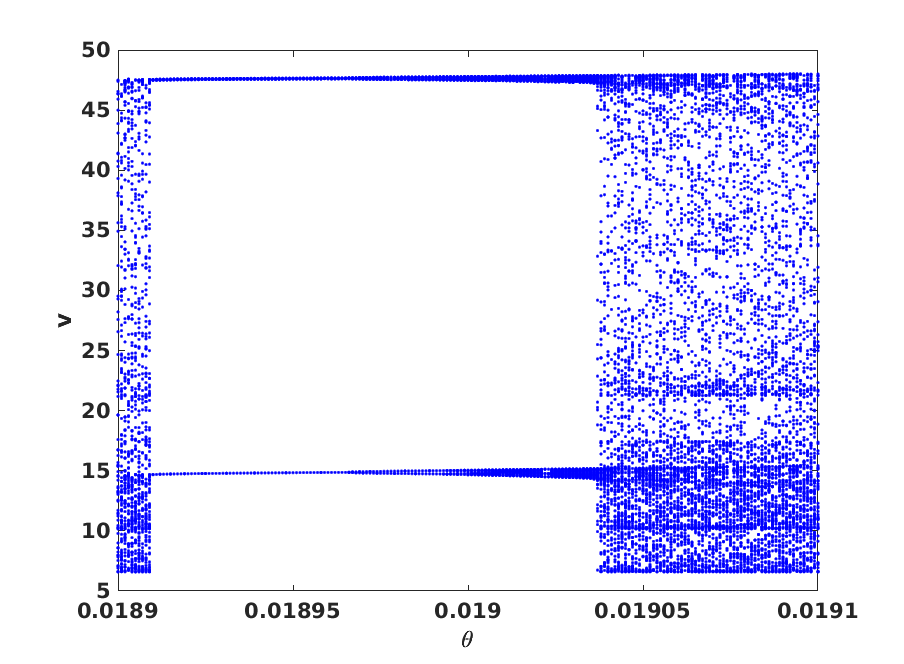}
    \caption{Bifurcation diagram of prey populations with respect to Allee parameter, in the range $\theta \in [0.0189 : 0.0191]$. Here we display the local maxima of the prey population. 
The parameter values in Eqn.~\ref{eq:finalmodel} are as mentioned in the text.}
    \label{zoom}
\end{figure}

However, interestingly, the system does have one parameter window where there is attractor widening and this gives rise to a markedly enhanced extreme event count. The peak observed in Fig.~\ref{fig:Dsit_no_extreme} can be directly correlated with a sudden attractor widening leading to a marked increase of extreme event in a narrow window of parameter space located near the crisis (see Fig.~\ref{zoom}). Additionally, for a narrow window around $\theta \sim 0.02$, the emergent dynamics is periodic. So the large deviations are no longer uncorrelated, and so they are not extreme events in the true sense.

Lastly we notice that the predator population shows extreme events for all values of $\theta \in [0,\theta_{c})$. So the predator population is most prone to experiencing unusually large deviations from the mean. We also observe that the probability of occurrence of extreme events in the predator population is not affected significantly by the Allee effect. This is in marked contrast to the case of vegetation and prey, where the Allee effect crucially influences the advent of extreme events. Also, for the predator population there is no marked transition from zero to finite $P_{ext}$ under increasing Allee parameter $\theta$, as evident for vegetation and prey populations.




\section{Effect of Noise in the System: Mitigation of Blow-Ups and Extreme Events}

Most realistic population models are not deterministic, as noise is ubiquitous in nature. So stochasticity must be incorporated into the models, since there are many external influences, such as migration, diversity and environmental fluctuation, present in the real ecosystem. For instance, in important earlier works the role of environmental fluctuation dependent fitness in population dynamics, namely  parametric noise, in extinction and persistence has been studied \cite{newref}.

In this work, we explore the interplay of stochasticity and extreme events, by investigating the system under the influence of additive noise.
It is of much relevance to explore if noise has any significant effect on the dynamics, for instance on the  unbounded vegetation growth under increasing Allee effect and on the emergent chaotic attractors. The other question of utmost interest is the following: does noise mitigate or aid the emergence of extreme events. This will be the focus of our investigation in this section.


Specifically we investigate the dynamics of the three species system \eqref{eq:finalmodel} under additive random noise $\xi(t)$, given by the following dynamical equations:

\begin{equation}
    \begin{split}
    \dot{u} &= f(u,v,w) + \xi_{1}(t),\\
    \dot{v} &= g(u,v,w) + \xi_{2}(t),\\
    \dot{w} &= h(u,v,w) + \xi_{3}(t),
    \end{split}
    \label{eq:model_with_additive_noise}
\end{equation}

\noindent where $f (u,v,w)$, $g (u,,w)$ and $h(u,v,w)$ have the functional forms given in Eq.~\ref{eq:finalmodel}, and $\xi_{i}(t),\,i\,=\,1,2,3$ are Gaussian white noises with zero mean and correlation function is given by $<\xi_{i}(t),\xi_{j}(t^{'})>\,=\,\sigma \delta(t-t^{'})\delta_{ij}$ for $i,j\,=\,1,2,3$. Here $\sigma$ represents the strength of the noise. To simulate the dynamics of this noise-driven system, we numerically solve the stochastic differential system by the explicit Euler-Maruyama scheme.

First we investigate if the boundedness of the system \eqref{eq:model_with_additive_noise} is affected by the presence of additive noise. Recall that the system, without noise, blows up as the magnitude of the Allee parameter $\theta$ increases beyond a threshold (cf. Fig.~\ref{fig:Prob_blowup}). Therefore it is important to examine if noise suppresses or enhances the probability of unbounded growth in the system under Allee effect.

\begin{figure}
    \centering
    \includegraphics[width=0.55\textwidth]{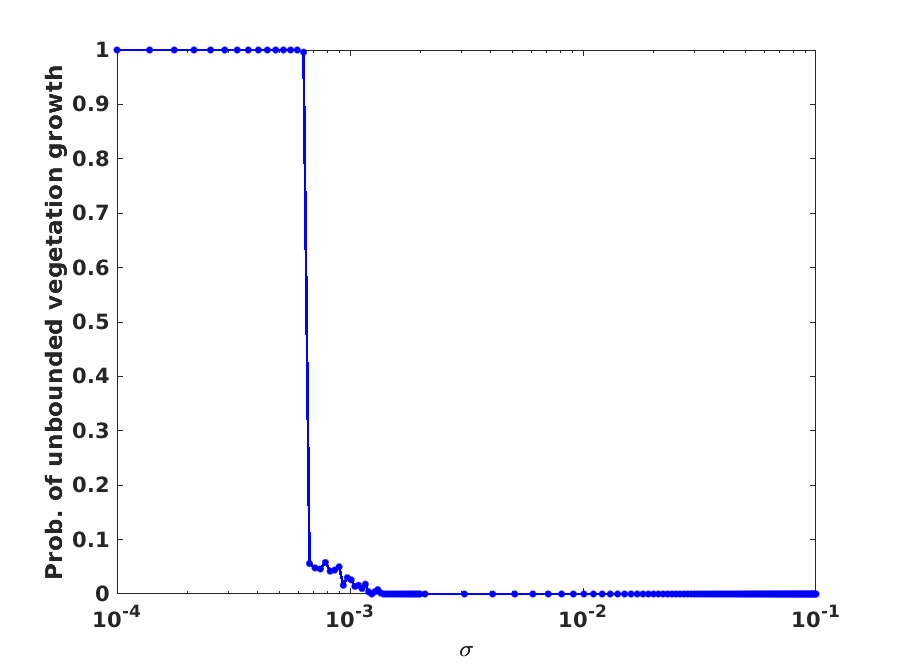}
    \caption{Probability of unbounded vegetation growth in presence of additive noise, with respect to noise strength $\sigma$. As in Fig.~\ref{fig:Prob_blowup}, a blow up to be considered to occur when the vegetation population exceeds $10^3$. Here the Allee parameter $\theta\,=\,0.1$, and the other system parameters are the same as in Fig.~\ref{fig:Prob_blowup}. The probability is estimated from a sample of $10^3$ initial states randomly distributed in a hyper-cube ($u \in [0:4], v \in [0:2], w \in [0:5]$) in phase space. \bigskip \bigskip \bigskip \bigskip \\}
    \label{fig:Prob_blowup_NS}
\end{figure}

Fig.~\ref{fig:Prob_blowup_NS} displays representative results for the probability of blow-ups in the population of vegetation, estimated for varying noise strengths $\sigma$, for the Allee parameter $\theta\,=\,0.1$. Note that the system without noise (i.e. $\sigma=0$) had significant probability of unbounded vegetation growth for this value of $\theta$ (see Fig.~\ref{fig:Prob_blowup}). It is clearly evident from the results in Fig.~\ref{fig:Prob_blowup_NS} that the probability of blow-ups for vegetation rapidly decreases to zero with increasing magnitude of noise strength. So the presence of noise helps to keep the populations bounded, indicating the {\em constructive role of noise in the stability of this three species system}. 

\begin{figure*}
    \centering
    \mbox{\subfigure[]{\includegraphics[width=0.45\textwidth]{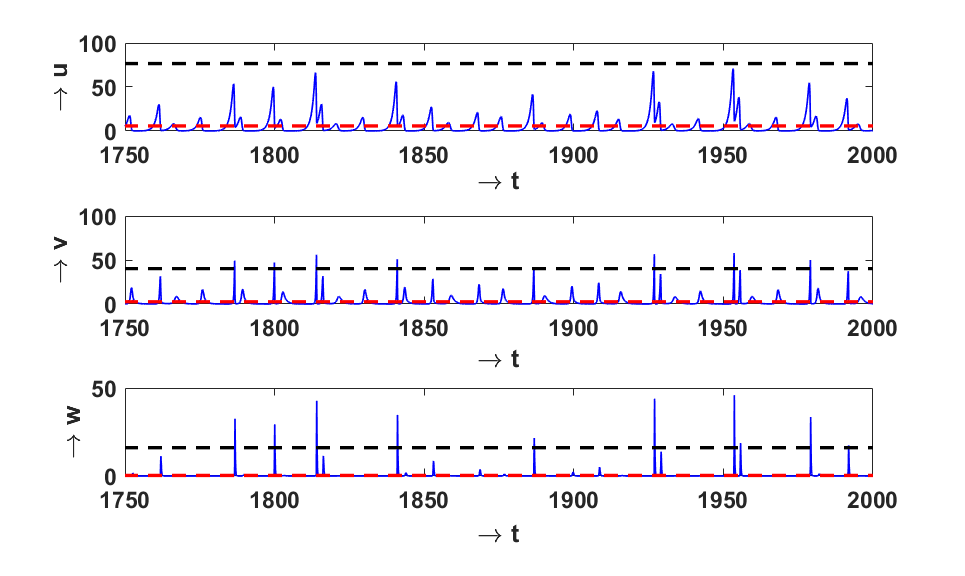}}
    \quad
    \subfigure[]{\includegraphics[width=0.45\textwidth]{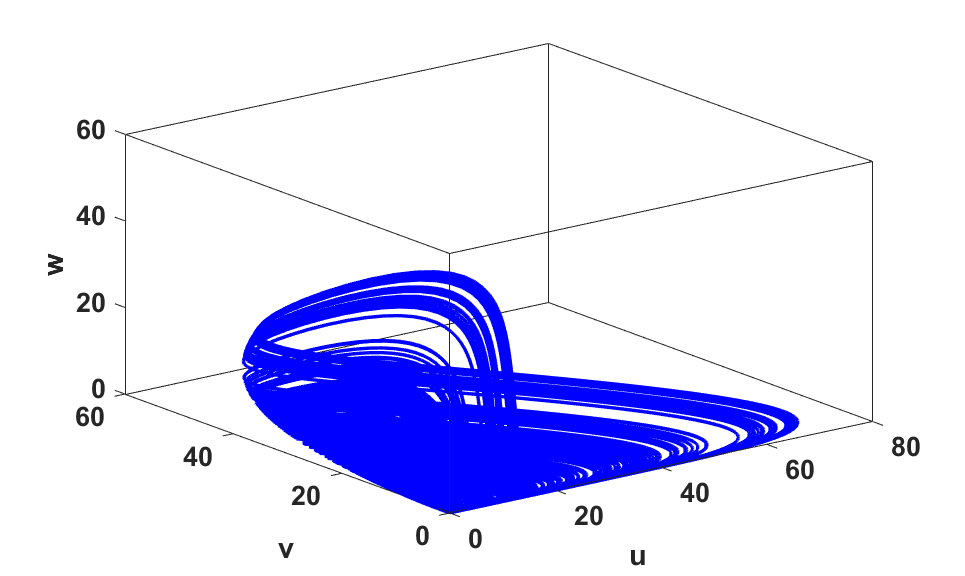}}}
    \mbox{\subfigure[]{\includegraphics[width=0.45\textwidth]{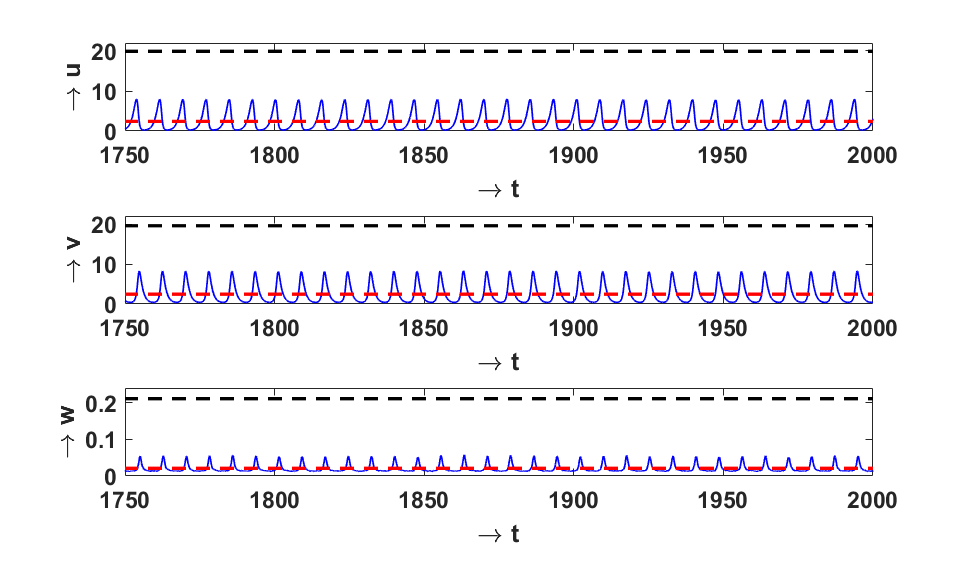}}
    \quad
    \subfigure[]{\includegraphics[width=0.45\textwidth]{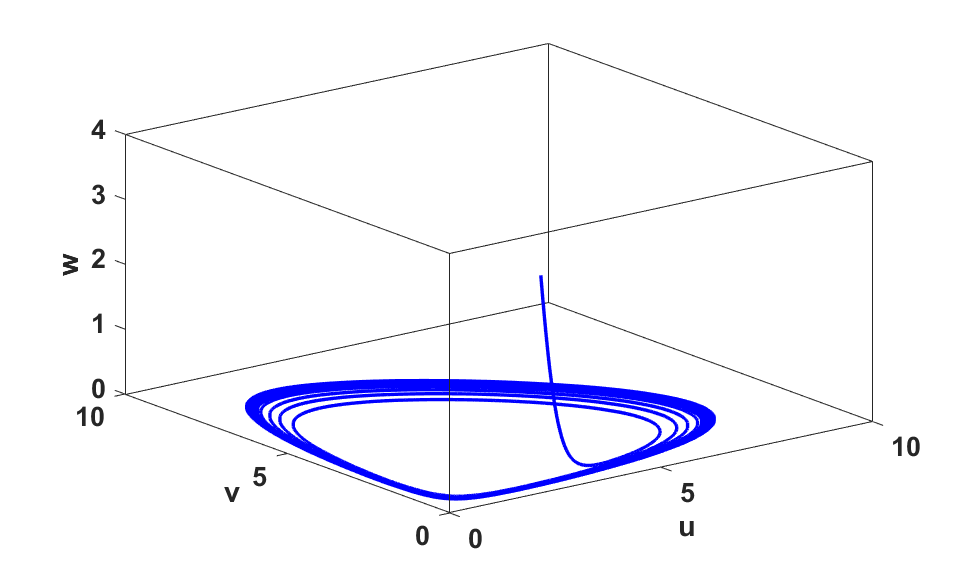}}
    }
    \mbox{\subfigure[]{\includegraphics[width=0.45\textwidth]{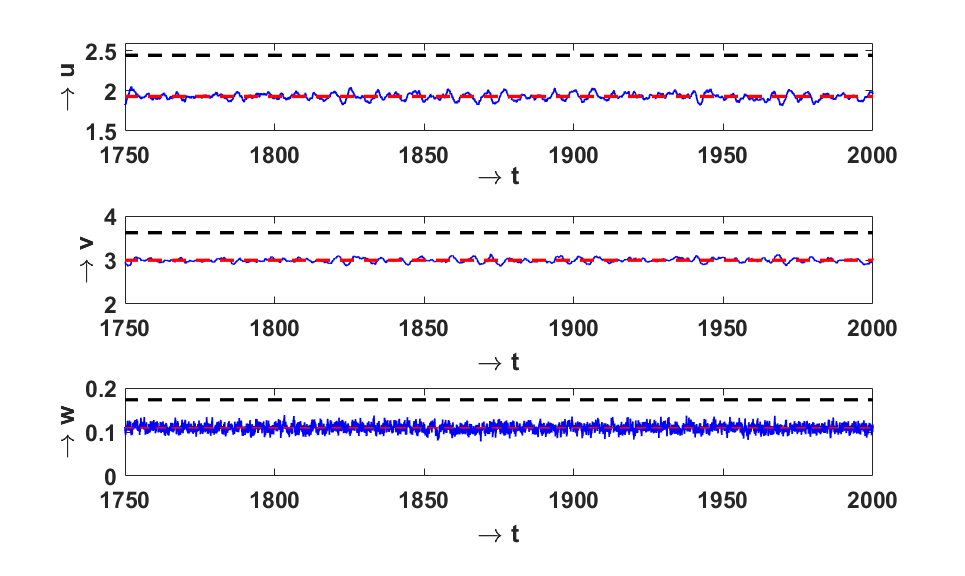}}
    \quad
    \subfigure[]{\includegraphics[width=0.45\textwidth]{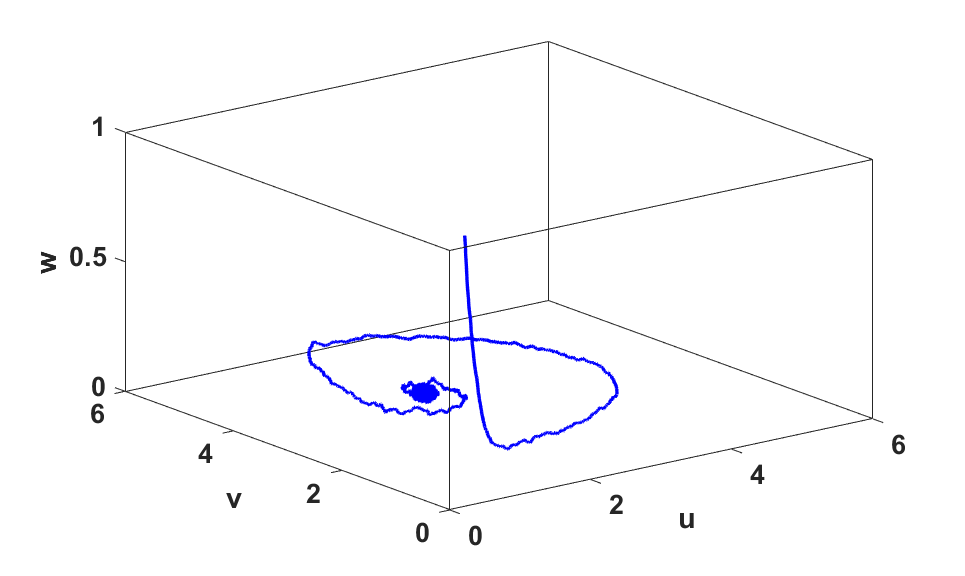}}}
    \caption{Time series and phase dynamics of the system \eqref{eq:model_with_additive_noise} with different $\sigma$. (a) $\sigma\,=\,10^{-4}$, (b) $\sigma\,=\,10^{-2}$ and (c) $\sigma\,=\,10^{-1}$. We keep all other parameters values are same as before except $\theta\,=\,0.024$.}
    \label{fig:TimeSeries and PhaseP_with_additive_Noise}
\end{figure*}

Next we examine how noise influences the extreme events which were observed to emerge in this system in the presence of Allee effect. We first examine the temporal evolution of the population densities and their corresponding phase-space attractors. In Fig.~\ref{fig:TimeSeries and PhaseP_with_additive_Noise} we display illustrative results of the times series and the phase-space attractors of the system governed by  Eqn.~\eqref{eq:model_with_additive_noise}, for different noise strength $\sigma$. It is observed that when the strength is very low ($\sigma \sim 10^{-4}$), all population densities fluctuate in an aperiodical manner and settle down to a chaotic attractor, as shown in Fig.~\ref{fig:TimeSeries and PhaseP_with_additive_Noise}a. Also extreme events occur in all populations for very low noise strengths, as is clear from the figure where the vegetation, prey and predator populations can be seen to cross the $\mu + 7\sigma$ threshold. However with increasing strength of noise, these extreme events disappear from all populations in the system. Additionally, the populations are seen to fluctuate in a more regular almost-periodic manner (see Fig.~\ref{fig:TimeSeries and PhaseP_with_additive_Noise}b). On further increase of the noise strength all populations settle down to a quasi-fixed state, as evident from Fig.~\ref{fig:TimeSeries and PhaseP_with_additive_Noise}c. This suggests that noise transforms the chaotic behaviour of the system to a noisy fixed point. Importantly, the very long time intervals we sampled did not yield a single extreme event. That is, under increased noise strengths there is no evidence of extreme events any more, in either the vegetation, prey or predator populations. 
Thus we arrive at the following important conclusion: {\em Noise leads to quasi-fixed (non-zero) populations and the suppression of extreme events in
this three species system.}

\begin{figure}
    \centering
    \includegraphics[width=0.6\textwidth]{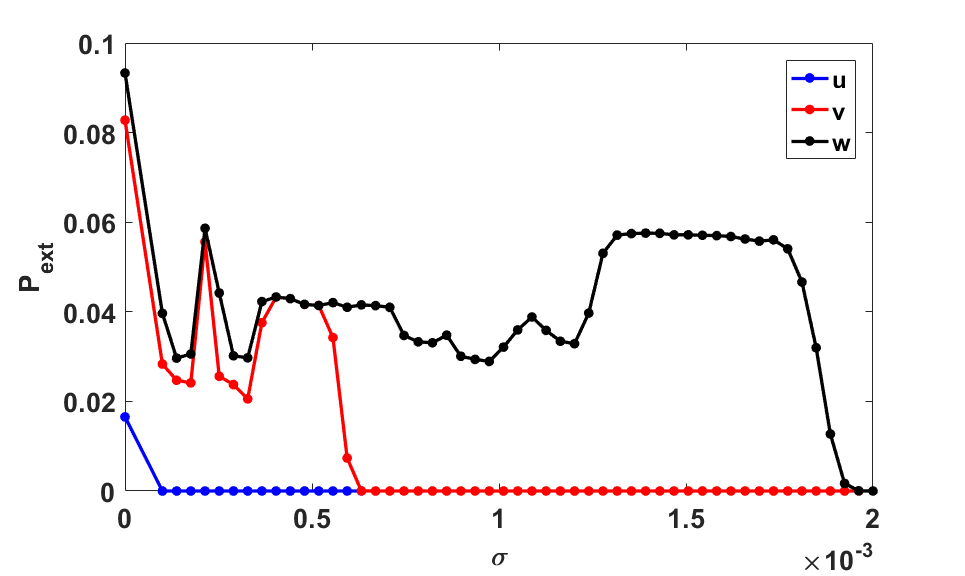}
    \caption{Probability of obtaining extreme events in unit time ($P_{ext}$) of the system \eqref{eq:model_with_additive_noise} with respect to the noise strength $\sigma$. Here we also consider a extreme event occurs when population level cross the $\mu + 7\sigma$ threshold. $P_{ext}$ for vegetation, prey and predator populations are shown by blue, red and black colour respectively. \bigskip \bigskip \bigskip \bigskip \\}
    \label{fig:Prob_extreme_wrt_NS}
\end{figure}

Further, in order to quantify how noise influences the emergence of extreme events, we again estimate the probability of obtaining extreme events, $P_{ext}$, under varying noise strength $\sigma$. The results are  exhibited in Fig.~\ref{fig:Prob_extreme_wrt_NS}. It is clear from the figure that the probability of obtaining extreme events is the lowest for vegetation and the highest for the predator population. This is consistent with the observations for the system without noise (see Fig.~\ref{fig:Dsit_no_extreme}). The new significant result here is that the probability of obtaining extreme events decreases to zero for increasing the noise strength $\sigma$. Therefore in presence of sufficiently strong additive noise, extreme events are suppressed in vegetation, prey and predator populations. This points to the novel finding that {\em stochasticity can lead to the mitigation of extreme events.}

\section{Discussion}

In summary, we explored the dynamics of a three-species trophic system incorporating the Allee Effect in the prey population. Our focus is on the emergence of extreme events in the system. In particular we address the significant question of whether or not Allee effect suppresses or enhances extreme events. 
Our key observations are as follows: First, under Allee effect the regular periodic dynamics changes to chaotic, as evident from the emergence of chaotic attractors for increasing Allee parameter $\theta$. Further, we find that the system exhibits unbounded growth in the vegetation population (a ``blow-up'') after a critical value of the Allee parameter. The most significant result is the observation of a critical Allee parameter beyond which the probability of obtaining extreme events becomes non-zero for all three population densities. Though the emergence of extreme events in the predator population is not affected much by the Allee effect, the prey population shows a sharp increase in the probability of obtaining extreme events after a threshold value of the Allee parameter $\theta$, and the vegetation population also yields extreme events for sufficiently strong Allee effect. An interesting open problem in this context would be to check the observation that the extreme events in the predator population are more pronounced than in prey and vegetation across other models, in order to establish the generality of this important trend in a larger class of models.

Lastly we consider the influence of additive noise on extreme events. First, we find that noise tames the unbounded vegetation growth induced by Allee effect. More interestingly, we demonstrate that stochasticity drastically diminishes the probability of extreme events in all three populations. In fact for sufficiently high noise, we do not observe any more extreme events in the system. This indicates that noise can mitigate extreme events, and has potentially important impact on the observability of extreme events in naturally occurring systems.\\
 





\bigskip
\bigskip
\bigskip




{\bf Conflict of Interest}\\

The authors declare that they have no conflict of interest.\\

{\bf Availability of data and material}\\

Data will be made available on reasonable request.\\



{\bf Authors' contributions}\\

SS conceived the problem, DS did all the simulations, DS and SS analyzed the results and wrote the manuscript together.\\

\bigskip
\bigskip
\bigskip
\bigskip

\bibliography{ref}

\end{document}